\documentclass[pdflatex,sn-mathphys,Numbered]{sn-jnl}

\usepackage{graphicx}%
\usepackage{multirow}%
\usepackage{amsmath,amssymb,amsfonts,bm}%
\usepackage{amsthm}%
\usepackage{mathrsfs}%
\usepackage[title]{appendix}%
\usepackage{xcolor}%
\usepackage{textcomp}%
\usepackage{manyfoot}%
\usepackage{booktabs}%
\usepackage{algorithm}%
\usepackage{algorithmicx}%
\usepackage{algpseudocode}%
\usepackage{listings}%

\newcommand{\Zp}{Z_\perp}
\newcommand{\Rey}{\mathrm{Re}}

\newcommand{\vinf}{\bm{v}_\infty}
\newcommand{\uinf}{\bm{u}_\infty}
\newcommand{\Vinf}{\bm{V}_\infty}
\newcommand{\ominf}{\bm{\omega}_\infty}
\newcommand{\Ominf}{\bm{\Omega}_\infty}

\newcommand{\Sinf}{\mathbb{S}_\infty}
\newcommand{\bbR}{\mathbb{R}}

\newcommand{\ex}{\bm{\hat{x}}}
\newcommand{\ey}{\bm{\hat{y}}}
\newcommand{\ez}{\bm{\hat{z}}}

\newcommand{\eX}{\bm{\hat{X}}}
\newcommand{\eY}{\bm{\hat{Y}}}
\newcommand{\eZ}{\bm{\hat{Z}}}

\newcommand{\er}{\bm{\hat{r}}}

\begin{document}

\title[Hydrodynamics of flow sensing in plankton]{Hydrodynamics of flow sensing in plankton}

\author*[1]{\fnm{Christophe} \sur{Eloy}}\email{celoy@centrale-med.fr}

\affil*[1]{\orgname{Aix-Marseille Univ}, \orgname{CNRS}, \orgname{Centrale Marseille}, \orgdiv{IRPHE}, \orgaddress{\city{Marseille}, \country{France}}}

\abstract{
Planktonic organisms, despite their passive drift in the ocean, exhibit complex responses to fluid flow, including escape behaviors and larval settlement detection. But what flow signals can they perceive? This paper addresses this question by considering an organism covered with sensitive cilia and immersed in a background flow. The organism is modeled as a spherical particle in Stokes flow, with cilia assumed to measure the local shear at the particle surface. This study reveals that, while these organisms can always measure certain components of the flow strain, bottom-heaviness is necessary to measure the horizontal component of vorticity. These findings shed light on flow sensing by plankton, contributing to a better understanding of their behavior.
}


\maketitle

Plankton play a crucial role in marine ecosystems \cite{Valiela1995}. By definition, they are composed of microscopic organisms that drift with the flow. But that doesn't mean that they are passive. Most of them are active and respond to flow signals \cite{Visser2001,Wheeler2019}. However, little is known about the signals they can measure. 

Among plankton, ciliates form a subgroup of unicellular organisms characterized by hair-like appendages known as cilia. Much research has delved into the locomotion of ciliates through the motion of these cilia \cite{Lighthill1952,Blake1971b}. 
However, eukaryotes also use cilia as a mechano-sensory system \cite{Machemer1985,Coste2010,Bezares-Calderon2020,Brette2021}. Different observations have shown how ciliates and other plankton use cilia to measure the flow \cite{Fuchs2013,Wheeler2015,Wheeler2019}, detect prey and predators \cite{Jakobsen2001,Chen2010,Kiorboe2014}, or find a larval settlement \cite{Butman1988,Harii2002}. 

In this paper, I will consider the hydrodynamics of flow sensing by a ciliated organism modeled as a passive spherical particle immersed in a Stokes flow. 

\section{Problem statement}\label{sec.pb_statement}

\begin{figure}[t]
\begin{center}
\includegraphics[scale=0.25]{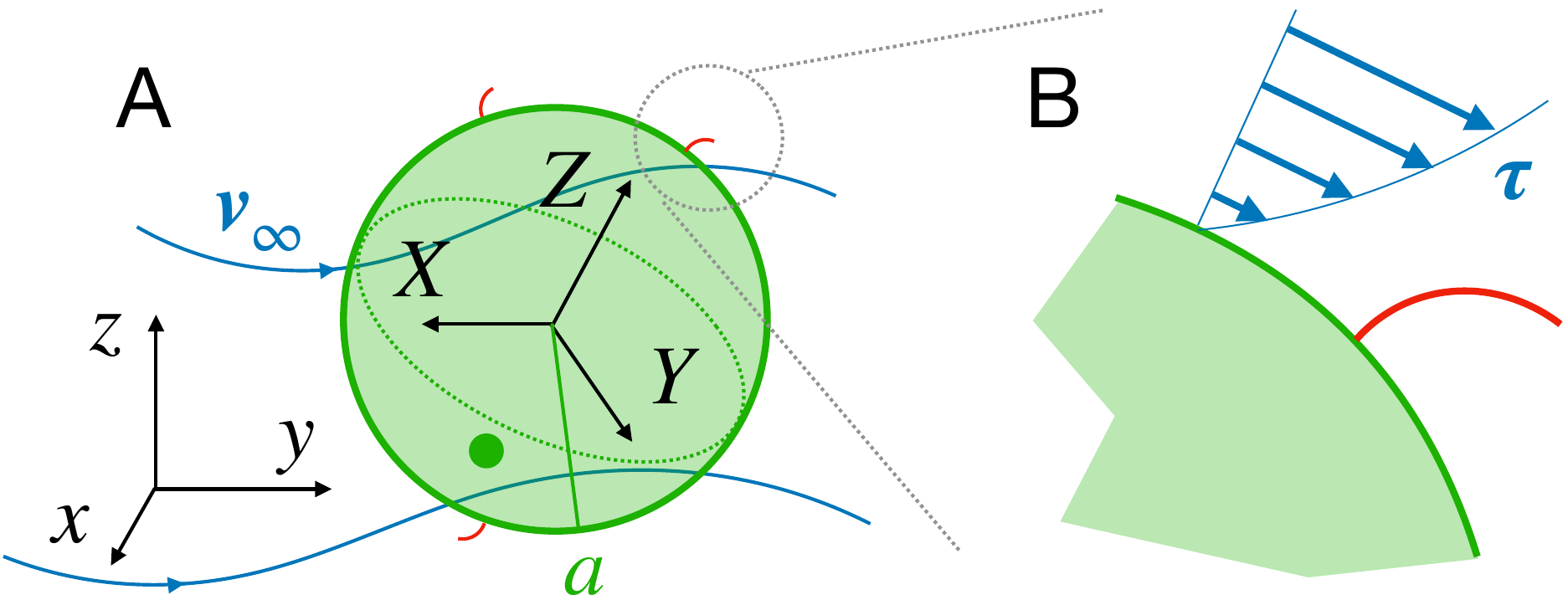}
\caption{
(A) Sketch of the problems and the notations used. A spherical particle of radius $a$ is immersed in a flow $\vinf$. 
(B) On its surface, mechano-sensitive cilia are deformed by local shear $\bm{\tau}$.}
\label{fig1}
\end{center}
\end{figure}

Let us consider a passive spherical particle of radius $a$ in a background flow (Fig.~\ref{fig1}A). We attach the framework $XYZ$ to this particle and we note $\eX$, $\eY$, $\eZ$ the associated unit vectors. We assume that the particle is not neutrally buoyant, its density being $\rho + \Delta\rho$, with $\rho$ the density of the surrounding fluid. 

The background flow is attached to the laboratory framework $xyz$. The change of framework from $xyz$ to $XYZ$ is given by the rotation matrix $\bbR$, such that $\bm{X} = \bbR\cdot\bm{x}$, where $\bm{x}$ is a vector in the framework $xyz$ and $\bm{X}$ the same vector in the framework $XYZ$. 

The background flow may be turbulent or generated by a predator. We note its smallest scale $\eta$ (the Kolmogorov scale if the flow is turbulent), its associated speed $v_\eta$, and its dynamic viscosity $\mu$. We assume that their values are such that the flow is smooth for the particle ($a\ll\eta$) and the Reynolds number is asymptotically small ($\Rey=\rho a v_\eta /\mu \ll 1$). The two assumptions are equivalent in a turbulent flow since $\Rey=a/\eta$ in that case. Typical values for the Kolmogorov scale in the ocean are of the order of $1\,$mm \cite{Fuchs2016}, meaning that our study applies to unicellular organisms, which are typically of size below $100\,\mu$m. 

Since the Reynolds number is asymptotically small, the flow obeys the Stokes equations for incompressible fluid
\begin{eqnarray}
-\nabla p + \mu \Delta \bm{u} &=& 0,\\
\nabla\cdot\bm{u} &=& 0,
\end{eqnarray}
with $\bm{u}$ the velocity field and $p$ the pressure. The total flow $\bm{u}$ can be decomposed into a sum of the background flow $\vinf$ and a perturbation $\bm{v}$ due to the presence of the particle. Locally, around the particle, the background flow $\vinf$ can be linearized
\begin{eqnarray}
\vinf(\bm{r}) 	& \approx & \Vinf + \nabla\vinf \cdot \bm{r},\\
				& \approx & \Vinf + \Ominf \times \bm{r} + \Sinf \cdot \bm{r} ,
\end{eqnarray}
where $\Vinf = \vinf(0)$ is the mean flow, $\Sinf = \frac{1}{2}\left(\nabla\vinf + \nabla\vinf^T\right)$ is the second-order rate-of-strain tensor (the symmetric part of the gradient tensor), and $\Ominf = \tfrac{1}{2}\ominf $ the angular velocity (associated with the antisymmetric part of the gradient tensor), with $\ominf = \nabla\times\vinf$ the vorticity . These quantities are all evaluated at the particle center $\bm{r}=0$, taken to the be the origin of the framework without loss of generality.

\subsection{Decomposition into two problems: gravity and strain}\label{sec.mobility}

We assume that the particle surface is partially covered by sensors that can perceive the local shear (Fig.~\ref{fig1}B). For ciliated unicellular organisms, these sensors are sensitive cilia. This shear is a vector field on the particle's spherical surface. The question now is to relate this field with the information on the background flow and the gravity orientation.  

We can decompose our general problem into two independent problems: 
\begin{enumerate}
\item Finding the flow $\bm{u}_g$ associated with the externally forced motion (here gravity) in a linear flow without strain $\vinf=\Vinf + \Ominf \times \bm{r}$;
\item Finding the flow $\uinf$ associated to a pure strain flow $\Sinf \cdot \bm{r}$ in the absence of external force. 
\end{enumerate}
By virtue of the linearity of Stokes equations, the total flow $\bm{u}$ solution of the general problem is
\begin{equation}\label{eq:decompose_v_alt}
\bm{u}  = \bm{u}_g + \uinf.
\end{equation}
Each of these terms is associated with a velocity shear on the sphere surface, which we aim to calculate. This is this shear that the organism is susceptible to sense. The total shear can be decomposed in the same manner as the velocity field 
\begin{equation}\label{eq:decompose_tau}
\bm{\tau}  	= \left.\frac{\partial \bm{u}_\parallel}{\partial r}\right|_{r=a}
			= \bm{\tau}_g + \bm{\tau}_\infty ,
\end{equation}
where $\bm{u}_\parallel$ is the projection onto the direction tangential to the spherical surface of the relative velocity field between the flow and the particle. 

\section{Continuous flow sensing}\label{sec.flow_sensing_continuous}

We shall now calculate the flow $\bm{u}$ and shear $\bm{\tau}$ associated to the two problems:  $\bm{u}_g$ and $\bm{\tau}_g$ for a buoyant particle in a linear flow with zero strain; $\bm{u}_\infty$ and $\bm{\tau}_\infty$ for a non-buoyant particle in a pure strain. In this section, we will consider that the particle has a perfect knowledge of the shear field $\bm{\tau}$ over its whole surface and we will examine how this information is related to the background flow and orientation of gravity. 

\subsection{First problem: shear due to gravity}\label{sec.flow_gravity}
\begin{figure}[t]
\begin{center}
\includegraphics[scale=0.25]{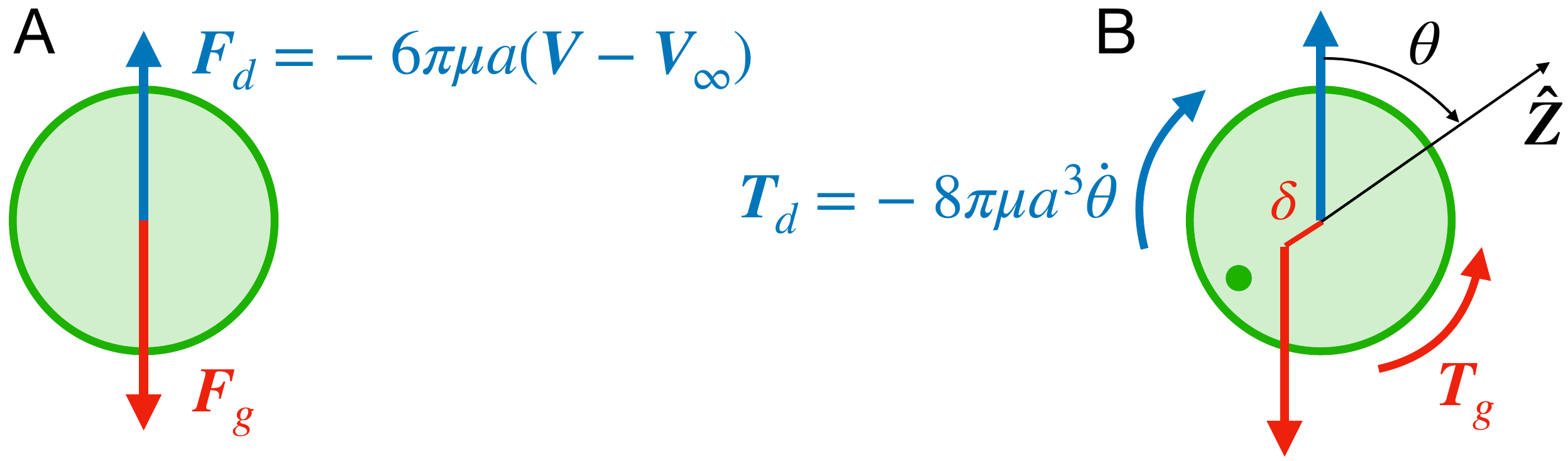}
\caption{
(A) The balance between the weight $\bm{F}_g$ and the viscous drag gives a settling velocity $V_\mathrm{sink}$. 
(B) The balance between the bottom-heavy torque $\bm{T}_g$ and the viscous torque gives a title angle.}
\label{fig2}
\end{center}
\end{figure}

We consider the first problem of a spherical particle in a linear flow $\Vinf + \Ominf \times \bm{r}$. The particle is subject to the gravitational force and torque
\begin{equation}
\bm{F}_g = \frac{4}{3}\pi a^3\Delta\rho \,\bm{g}, \quad
\bm{T}_g = \frac{4}{3}\pi a^3\rho\, \bm{\delta} \times \bm{g},
\end{equation}
with $\bm{g}$ the acceleration of the gravity. The torque $\bm{T}_g$ is the bottom-heavy torque due to the offset $\bm{\delta}$  between the center of gravity and the center of buoyancy (Fig.~\ref{fig2}). 

\begin{figure}[t]
\begin{center}
\includegraphics[scale=0.25]{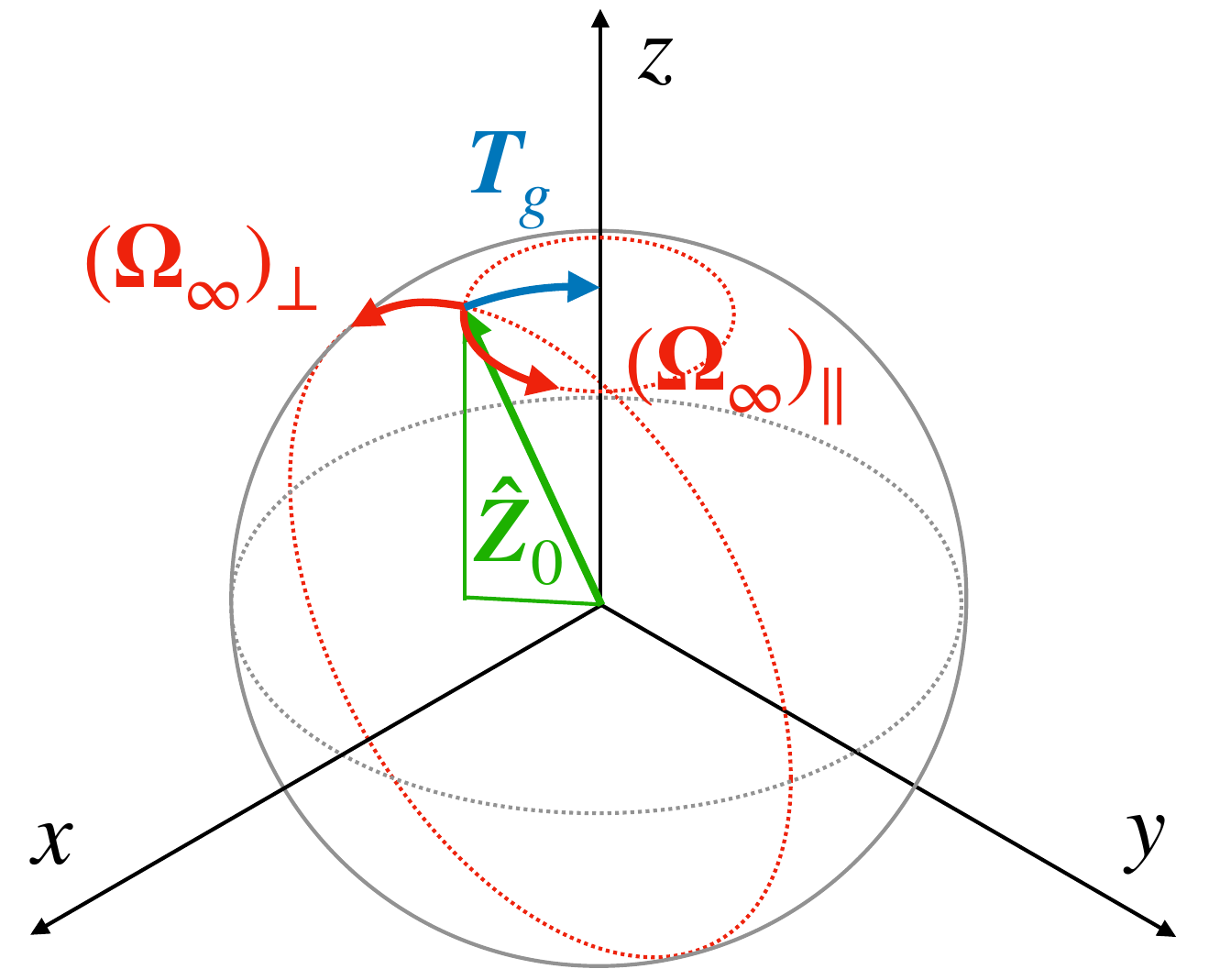}
\caption{
Sketch showing that there exists an equilibrium position $\bm{Z}_0$ when the particle is immersed in a rotation rate $\Ominf$ that includes a horizontal component $(\Ominf)_\perp$ along $x$ and a vertical component $(\Ominf)_\parallel$ along $z$. The associated viscous torques (in red) are compensated by the bottom-heavy torque $\bm{T}_g$ (in blue). }
\label{fig3}
\end{center}
\end{figure}

In the Stokes limit, the particle is force- and torque-free. This means that the gravitational force $\bm{F}_g$ is compensated by a viscous drag force, which is proportional to the relative velocity between the particle and the fluid. Similarly, the gravitational torque $\bm{T}_g$ is compensated by a viscous torque proportional to the relative angular velocity between the particle and the flow. It yields the following mobility equations \cite{Kim1991}
\begin{eqnarray}
\bm{V} - \Vinf 		&=& \frac{1}{6 \pi \mu a} \bm{F}_g = - V_\mathrm{sink} \,\ez, \\
\bm{\Omega} - \Ominf &=& \frac{1}{8 \pi \mu a^3} \bm{T}_g = \frac{1}{2 \tau}  \, \eZ \times \ez, \label{eq.rotation}
\end{eqnarray}
with $\bm{V}$ and $\bm{\Omega}$ the translational and angular velocities of the particle, $V_\mathrm{sink} = 2 a^2\Delta\rho g/(9 \mu)$ the sinking velocity, and $\tau=3 \mu/(\rho \delta g)$ the typical bottom-heavy tilting time, with $\bm{\delta} = \delta \eZ$ the offset. Equation \eqref{eq.rotation} can be rewritten as a first-order differential equation for $\eZ$
\begin{equation}
\frac{d \eZ}{d t} = \bm{\Omega} \times \eZ = \left( \Ominf + \frac{1}{2 \tau}  \, \eZ \times \ez\right) \times \eZ.
\end{equation}
This equation admits a fixed point $\eZ_0$ (Fig.~\ref{fig3}), that satisfies
\begin{subequations}\label{eq.Z0}
\begin{eqnarray}
\eZ_0 \cdot \ey &=& 
	-\frac{1+\alpha^2-\sqrt{(\alpha^2+1)^2-4 \alpha_\perp^2}}{2 \alpha_\perp} = -\alpha_\perp + O\left(\alpha^3\right),\\ 
\eZ_0 \cdot \ez &=& \sqrt{
	\frac{1-\alpha^2+\sqrt{(\alpha^2+1)^2-4 \alpha_\perp^2}}{2}} = 1 - \frac{\alpha_\perp^2}{2} + O\left(\alpha^4\right),
\end{eqnarray}
\end{subequations}
with $\ex$ aligned with the horizontal component of the rotation vector $(\Ominf)_\perp = (\mathbb{I} - \ez\ez) \cdot \Ominf$ (without loss of generality), $\alpha=2\tau\|\Ominf\|$, and $\alpha_\perp=2\tau \| (\Ominf)_\perp \| \le \alpha$.
The third component of $\eZ_0$ can be found by enforcing $\|\eZ_0 \|=1$, with $\eZ_0 \cdot \ex$ of same sign as $\Ominf\cdot \ez$. 

The flow velocity corresponding to the translational and angular velocities, $\bm{V}$ and $\bm{\Omega}$, is a classical result \cite{Guazzelli2011}: the rotation of the sphere is associated with a rotlet and the translation with a combination of a stokeslet and a doublet. 
The resulting flow is $\bm{u} = \bm{u}_V + \bm{u}_\Omega$ with 
\begin{eqnarray}
\bm{u}_V & = &	\Vinf + \frac{3}{4}\left(\frac{a}{r^3} - \frac{a^3}{r^5}\right)\er\er \cdot (\bm{V} - \Vinf) + 
			\left(\frac{3a}{4r} +  \frac{a^3}{4r^3}\right) (\bm{V} - \Vinf), \\
\bm{u}_\Omega & = &	r \,\Ominf \times \er +
			\frac{a^3}{r^3} (\bm{\Omega} - \Ominf) \times \er,
\end{eqnarray}
which corresponds to tangential velocities relative to the particle respectively of
\begin{eqnarray}
\left(\bm{u}_V\right)_\parallel &=&
 	\left(\frac{3a}{4r} + \frac{a^3}{4r^3} - 1\right) \left( \mathbb{I} - \er\er \right) \cdot (\bm{V} - \Vinf),\\
\left(\bm{u}_\Omega\right)_\parallel &=&
			\left(\frac{a^3}{r^2} - r\right) (\bm{\Omega} - \Ominf) \times \er,
\end{eqnarray}
with $(\bm{u}_g)_\parallel = (\bm{u}_V)_\parallel +(\bm{u}_\Omega)_\parallel$ and $\bm{r} = r \er$. We can now calculate the associated shear 
\begin{eqnarray}
\bm{\tau}_g 	& = & \left.\frac{\partial \left(\bm{u}_g\right)_\parallel}{\partial r}\right|_{r=a}, \\
 				& = & -\frac{3}{2a} \left( \mathbb{I} - \er\er \right) \cdot(\bm{V}-\Vinf)  - 
				3(\bm{\Omega} - \Ominf) \times \er ,\\
 				& = & \frac{a\Delta\rho g}{3 \mu}  \left( \mathbb{I} - \er\er \right) \cdot \ez + 
				\frac{3}{2 \tau} \left(\ez \times \eZ_0\right) \times \er . \label{eq.taug}
\end{eqnarray}
Here, the first term corresponds to the sinking of the particle, and the second term to the bottom-heaviness.  
The sinking component of the shear is independent of the background flow, as expected. Sensing this shear only informs the particle on the orientation $\ez$ of the vertical.
The bottom-heaviness component of the shear corresponds to a rotation vector proportional to $\ez \times \eZ_0$. In the general case, this information is not easy to link to the flow because $\eZ_0 \cdot \ez$, given by Eq. (\ref{eq.Z0}b),  is a nonlinear combination of the different components of $\Ominf$. The particular case of rapid bottom-heavy tilting compared to the typical time of the flow $1/\|\Ominf\|$ corresponds to the limit $\alpha \ll 1$ in Eqs (\ref{eq.Z0}a,b). In that case $\ez \times \eZ_0 \approx 2 \tau (\Ominf)_\perp$ and the bottom-heavy shear allows the particle to directly sense the horizontal component of the vorticity. 

For a non-buoyant bottom-heavy particle with rapid tilting time, the information about the shear $\bm{\tau}_g$ due to gravity allows the particle to measure $\ez$ the vertical direction and $(\Ominf)_\perp$ the horizontal component of the background flow rotation rate (i.e., twice the vorticity). 

\subsection{Second problem: shear due to strain}\label{sec.flow_strain}

The second problem is to determine the flow and the associated shear when the spherical particle is immersed in a pure strain flow $\Sinf \cdot \bm{r}$. The solution of the Stokes equations with boundary condition $\bm{v}  = - \Sinf \cdot \bm{r}$ in $|\bm{r}|=a$ is classical \cite{Lauga2016}  
\begin{equation}
\bm{v} = 	- \frac{a^5}{r^5} \Sinf \cdot \bm{x} 
			- \frac{5}{2}\left(\frac{a^3}{r^5}-\frac{a^5}{r^7}\right) 
				\left( \bm{x} \cdot \Sinf \cdot \bm{x}\right) \bm{x},
\end{equation}
with the associated pressure $p = 5\mu a^3  \left( \bm{x} \cdot \Sinf \cdot \bm{x}\right) /r^5$.
The total flow is
\begin{equation}
\uinf = 	\left(r - \frac{a^5}{r^4}\right) \Sinf \cdot \er 
			- \frac{5}{2}\left(\frac{a^3}{r^2}-\frac{a^5}{r^4}\right) 
				\left( \er \cdot \Sinf \cdot \er\right) \er,
\end{equation}
and the associated shear is
\begin{eqnarray}
\bm{\tau}_\infty 	& = & \left.\frac{\partial \uinf}{\partial r}\right|_{r=a}, \\
 				& = & 5\, \Sinf \cdot \er - 5 \left( \er \cdot \Sinf \cdot \er\right) \er,\\
				& = & 5 \left( \mathbb{I} - \er\er \right) \cdot \Sinf \cdot \er, \label{eq.tau_S}
\end{eqnarray}
which is simply (up to a factor 5), the projection onto the sphere surface of the pure strain $\Sinf$. So, in principle, knowing the shear field $\bm{\tau}_\infty$, the particle has information on the 5 independent components of the trace-free symmetric second-order tensor $\Sinf$. 

\section{Discrete and noisy flow sensing}\label{sec.flow_sensing_discrete}

So far, we have examined the information about the flow that a particle can gather by measuring the complete shear field on its surface. In practice, however, there is a finite number of sensors, and their measurements may be noisy. We shall now examine this case of discrete and noisy flow sensing. 

\subsection{Inverse problem}\label{sec.inverse_pb}

We assume that there is a linear relationship between the total shear $\bm{\tau}$ and an information vector $\bm{X}$. In general, the vector $\bm{X}$ has 10 dimensions: the 5 independent components of $\Sinf$, the orientation of gravity with respect to the particle orientation (2 components), and the horizontal component of the vorticity (3 components).  Formally, this can be written as
\begin{eqnarray}
\tau_\theta(\theta, \phi)	& = & \bm{M}_\theta (\theta, \phi) \cdot \bm{X}, \label{eq:sensingtheta}\\
\tau_\phi  (\theta, \phi)	& = & \bm{M}_\phi   (\theta, \phi) \cdot \bm{X}, \label{eq:sensingphi}
\end{eqnarray}
where $\tau_\theta$ and $\tau_\phi$ represents the two components of the shear in the $(r,\theta,\phi)$ spherical coordinate system of the particle attached to $XYZ$. Note that the particle measures the strain tensor $\mathbb{S}_{XYZ}$ and the horizontal rotation rate $\Omega_{XYZ}$ in its framework $XYZ$. These quantities can be obtained from the corresponding tensor $\Sinf$ and horizontal component of the vector $(\Ominf)_\perp$ in the framework $xyz$ using the rotation matrix $\mathbb{R}$ 
\begin{equation}
\mathbb{S}_{XYZ} = \mathbb{R} \cdot \Sinf \cdot \mathbb{R}^T, \quad
\bm{\Omega}_{XYZ} = \mathbb{R} \cdot \left( \Ominf \right)_\perp,
\end{equation}
where $\mathbb{R}$ can conveniently be written using the coordinates $[Z_x, Z_y, Z_z]$ of $\eZ$ in the framework $xyz$
\begin{equation}
\mathbb{R} = \left(\begin{array}{ccc}
-Z_y/\Zp 	&  Z_x/\Zp 		& 0 \\
-Z_xZ_z/\Zp & -Z_yZ_z/\Zp 	& \Zp \\
 Z_x 		&  Z_y 			& Z_z
\end{array}\right), \quad \mbox{with } \Zp=\sqrt{1-Z_z^2}.
\end{equation}

When there is a discrete number of sensors, the sensing information can be gathered in a vector $\bm{s}$ such that
\begin{equation}
\bm{s} = [\tau_*(\theta_1,\phi_1), \cdots \tau_*(\theta_m,\phi_m)],
\end{equation}
where $\tau_*$ represent either $\tau_\theta$ or $\tau_\phi$ and $(a,\theta_i,\phi_i)$ represents the spherical coordinates of the $i$th sensor.  Using \eqref{eq:sensingtheta} and \eqref{eq:sensingphi}, we can thus write the inverse problem as 
\begin{equation}\label{eq:inverseproblem}
\bm{s} = \mathbb{M} \cdot \bm{X} + \bm{\xi}, 
\end{equation}
where $\bm{\xi}$ represents an additive noise and $\mathbb{M}$ is the $m\times n$ tensor obtained by assembling the $m$ vectors $\bm{M}_*(\theta_i,\phi_i)$ of length $n=10$. The goal is to find the best approximation of $\bm{X}$ given $\bm{s}$. 

\subsection{Singular value decomposition}\label{sec.SVD}

To solve the inverse problem \eqref{eq:inverseproblem}, we  use the singular value decomposition of $\mathbb{M}$, which can be written as
\begin{equation}
\mathbb{M} = \mathbb{U} \cdot \mathbb{D} \cdot \mathbb{V}^T,
\end{equation}
where 
\begin{itemize}
\item $\mathbb{U}$ is a $m\times m$ matrix, its columns $\bm{u}_i$ being the eigenvectors of $\mathbb{M}\cdot\mathbb{M}^T$
\item $\mathbb{D}$ is a $m\times n$ rectangular diagonal matrix with, on its diagonal, $\lambda_i$ the square roots of the eigenvalues of $\mathbb{M}^T\cdot\mathbb{M}$ 
\item $\mathbb{V}$ is a $n\times n$ matrix, its column $\bm{v}_i$ being the eigenvectors of $\mathbb{M}^T\cdot\mathbb{M}$
\end{itemize}
An alternative notation is 
\begin{equation}\label{eq:SVD}
\mathbb{M} = \sum_{i=1}^m \lambda_i  \bm{u}_i \bm{v}_i^T,
\end{equation}
where $\lambda_i$ are the positive singular values of $\mathbb{M}$ in  decreasing order, $\bm{u}_i$ are the left-singular vector in the sensing space, and $\bm{v}_i$ are the right-singular vector in the flow information space of $\bm{X}$.

Coming back to the inverse problem \eqref{eq:inverseproblem}, the decomposition \eqref{eq:SVD} can be used to write the pseudo inverse of $\mathbb{M}$ such that, if $\bm{y}$ was the sensing information without noise
\begin{equation}\label{eq.M}
\bm{y} = \mathbb{M} \cdot \bm{X}, 
\end{equation}
then best approximation of $\bm{X}$ could be written as
\begin{equation}
\bm{X}^+ = \sum_{i=1}^{\mathrm{rank}\;\mathbb{M}} 
	\left( \frac{\bm{u}_i^T\cdot \bm{y}}{\lambda_i} \right) \cdot  \bm{v}_i,
\end{equation}
where $\bm{X}^+$ is a vector in the vectorial space of dimension rank $\mathbb{M}$, whose basis is formed of the first $\bm{v}_i$ vectors (for rank $\mathbb{M} \leq n$). If rank $\mathbb{M} > n$, then $\bm{X}^+$ represents the best least-square approximation of $\bm{X}$.  

When we use the same pseudo-inverse with noise, we see that we obtain an approximate $\bm{X}$, noted $\bm{\tilde{X}}$, that can be written as
\begin{equation}
\bm{\tilde{X}} = \sum_{i=1}^{\mathrm{rank}\;\mathbb{M}} 
	\left( \frac{\bm{u}_i^T\cdot \bm{s}}{\lambda_i} \right) \cdot  \bm{v}_i
	= \bm{X}^+ + \sum_{i=1}^{\mathrm{rank}\;\mathbb{M}} 
	\left( \frac{\bm{u}_i^T\cdot \bm{\xi}}{\lambda_i} \right) \cdot  \bm{v}_i.
\end{equation}
From this expression, we see that $1/\lambda_i$ gives an idea of how the noise $\bm{\xi}$ associated to a given eigenvector $\bm{v}_i$ may be amplified when trying to inverse the problem. This is why it is best to concentrate on the singular modes for which $\lambda_i$ is over a certain threshold $\lambda_c$. 

\subsection{Example 1: buoyant particle with 2 sensors}\label{sec.example1}

Let us consider a simple case with two sensors located in $\er=+\eZ$ and $\er=-\eZ$. We assume that the particle is buoyant and that its bottom-heaviness is strong such that the tilting time is short compared to the background flow vorticity [$\alpha \ll 1$ in Eqs (\ref{eq.Z0}a,b)]. In the XYZ framework, the particle will measure two shear vectors $\bm{\tau}_+$ and $\bm{\tau}_-$. Using Eqs \eqref{eq.taug} and \eqref{eq.tau_S}, it is easy to show that these vectors satisfy
\begin{eqnarray}
\bm{\tau}_+ + \bm{\tau}_- &=&  -\frac{2a\Delta\rho}{3 \mu}\left(g_X \eX + g_Y \eY\right),\\
\bm{\tau}_+ - \bm{\tau}_- &=&  
	\left( 
		 10 S_{XZ} + 6 \Omega_Y
	\right) \eX +
	\left( 
		 10 S_{YZ} - 6 \Omega_X
	\right) \eY ,
\end{eqnarray}
such that the sum of the two signals informs about the direction of gravity $\bm{g}$ and the difference informs some components of the strain $\mathbb{S}_{XYZ}$ and the rotation rate $\bm{\Omega}_{XYZ}$. Note that the gradient tensor $\mathbb{G}$ of the background flow field has two components that share the same ones as those measured by the difference of shears: $G_{XZ}=S_{XZ} + \Omega_Y$ and $G_{YZ}=S_{YZ} - \Omega_X$.

\subsection{Example 2: non-buoyant particle with 4 sensors}\label{sec.example2}

We now consider the case of 4 sensors measuring only the polar component of the shear $\tau_\theta$, these sensors being regularly arranged along an circle of constant polar angle $\theta_0$ ($\theta_0 \in [0, \frac{\pi}{2}]$ without loss of generality)
\begin{equation}
\theta_i = \theta_0, \quad \phi_i = \frac{i-1}{n}2 \pi, 
\end{equation}
with $n=4$.

For simplicity, we also consider that the particle is neutrally buoyant and has no bottomheaviness. In that case, the particle can only measure the strain and we reduce the unknown $\bm{X}$ to a 5-component vector 
\begin{equation}
\bm{X} = [S_{ZZ}, S_{XX} - S_{YY}, S_{XY}, S_{XZ}, S_{YZ}].
\end{equation}
Using the expression of $\bm{\tau}_\infty$ given by Eq. \eqref{eq.tau_S}, the matrix $\mathbb{M}$ defined in Eq. \eqref{eq.M} can be written as
\begin{equation}
\mathbb{M} = 
\left(\begin{array}{ccccc}
-\frac{15}{4}\sin2\theta_0 & \frac{5}{4}\sin2\theta_0 & 0 & 5\cos2\theta_0 & 0  \\[2pt]
-\frac{15}{4}\sin2\theta_0 &-\frac{5}{4}\sin2\theta_0 & 0 & 0 & 5\cos2\theta_0  \\[2pt]
-\frac{15}{4}\sin2\theta_0 & \frac{5}{4}\sin2\theta_0 & 0 &-5\cos2\theta_0 & 0 \\[2pt]
-\frac{15}{4}\sin2\theta_0 &-\frac{5}{4}\sin2\theta_0 & 0 & 0 &-5\cos2\theta_0 
\end{array}\right).
\end{equation}

Performing a singular value decomposition of this matrix, we find four singular values and four singular vectors:
\begin{eqnarray}
\lambda_1 & =  \frac{15}{2}\sin 2\theta_0, \quad & v_1 = S_{ZZ}, \\
\lambda_2 & =  \frac{ 5}{2}\sin 2\theta_0, \quad & v_2 = S_{XX} - S_{YY}, \\
\lambda_3 & = 5\sqrt{2}|\cos 2\theta_0|, \quad & v_3 = S_{XZ}, \\	
\lambda_4 & = 5\sqrt{2}|\cos 2\theta_0|, \quad & v_3 = S_{YZ}.		
\end{eqnarray}
We see that, depending on the value of $\theta_0$, the order of the singular values is different, but in general the four sensors allow to recover all independent components of the strain execept $S_{XY}$. 

For sensors near the pole or near the equator, such that $\sin 2\theta_0$ is small, the organism is only sensitive to the third and fourth singular vectors $v_3$ and $v_4$, which are the off-diagonal terms of the strain.  On the contrary, if the sensors are far from the pole and the equator such that $\theta_0=\frac{\pi}{4}$, the above expressions simplify into 
\begin{eqnarray}
\lambda_1 & =  \frac{15}{2}, \quad & v_1 = S_{ZZ}, \\
\lambda_2 & =  \frac{ 5}{2}, \quad & v_2 = S_{XX} - S_{YY},\end{eqnarray}
which means that the organism will be only sensitive to the diagonal terms of the strain. When we add noise, the meaningful signal will be dominated by $v_1 = S_{ZZ}$. The other component $v_2 = S_{XX} - S_{YY}$ will be likely harder to distinguish from noise, its singular value being 3 times smaller.

\section{Discussion}\label{sec.discussion}

In this paper, I showed how a planktonic organism can use flow sensing to measure certain components of the flow gradient. For simplicity, I modeled the organism as a spherical particle equipped with small sensitive hairs on its surface, capable of measuring the local shear without perturbing the flow. A sufficient number of such sensitive hairs enables the organism to measure the flow strain (the symmetric component of the flow gradient). However, to measure the flow vorticity (the antisymmetric component), the organism must be bottom-heavy.  Even with bottom-heaviness, the organism can only measure the horizontal component of the vorticity when the bottom-heavy tilting time is short compared to the vorticity timescale.  

Extending this study to swimming organisms is possible. For that, we need to account for the flow induced by the swimming motion. Subtracting the associated shear from the measurements allows for a direct comparison with the passive organism scenario, demonstrating the universality of the findings.

The potential extension of this research to organisms with ellipsoidal shapes introduces complexities in the mobility equations as shown by Jeffery \cite{Jeffery1922}. Unlike spherical shapes, strain components of the background flow influence the rotation of ellipsoidal particles, rendering it more challenging to divide the general problem into distinct problems. Moreover, the expression for shear flow at the surface of an ellipsoidal shape is notably more intricate.

In presenting this work, the primary goal was to better understand the organism's capabilities in flow sensing, outlining what it can and cannot measure. I leave for future studies to explore how organisms could optimize their behavior based on this information \cite{Monthiller2022}.

\bmhead{Acknowledgments}
This project has received funding from the European Research Council (ERC) under the European Union's Horizon 2020 research and innovation program (grant agreement No 834238).

\section*{Declarations}
Data sets generated during the current study are available from the corresponding author on reasonable request.

\bibliography{../../biblio.bib}

\end{document}